\documentclass[12pt]{iopart}
\usepackage{iopams} 
\usepackage{graphicx,epsf} 

\newcommand{\s}{\sigma}
\newcommand{\ep}{\epsilon}
\newcommand{\dl}{\delta}
\newcommand{\g}{\gamma}

\newcommand{\om}{\omega}
\newcommand{\pd}{\partial}
\newcommand{\bra}{\langle}
\newcommand{\ke}{\rangle}
\newcommand {\beq}{\begin{equation}}
\newcommand{\eeq}{\end{equation}}
\newcommand {\bseq}{\begin{subequations}}
\newcommand{\eseq}{\end{subequations}}
\newcommand{\bal}{\begin{align}}
\newcommand{\eal}{\end{align}}

\eqnobysec

\begin{document}

\title{ Exact spectral dimension of the random surface}

\author{I S Goncharenko}

\address{School of Natural Sciences, University of California, Merced, CA 95343}
\ead{igoncharenko@ucmerced.edu}
\begin{abstract}
We propose a new method of the analytical computation of the spectral dimension which is based on the equivalence of the random walk and the q-state Potts model with non-zero magnetic field in the limit $q\to 0$. Calculating the critical exponent of the magnetization $\dl$ of this model on the dynamically triangulated random surface by means of a matrix model technique we obtain that the spectral dimension of this surface is equal to two. 
\end{abstract}

\maketitle

\section{Introduction}

A diffusion in the dynamic medium  is an important problem due to its wealth of applications in many branches of physics. Examples include the diffusion through fluid membranes \cite{krolb, krola}, the diffusion in the presence of two-dimensional quantum gravity \cite{ABNRW, duplrw} and others. The main characteristic of the diffusion (random walk) is the probability that the particle returns to the starting site at time $t$. At large times one expects this probability to decay as $t^{-d_s/2},$ where $d_s$ is the spectral dimension of the underlying geometry. In the case of  ${\bf Z}^d$ we have $d_s=d$, which simply gives the dimension of the regular lattice. It is interesting how the fractal structure of  random lattices affects the spectral dimension. Despite the fact that the spectral dimension of many random graphs has been calculated analytically, in such cases, as branched polymers or generic trees $d_s=4/3$ \cite{wheabp, wheatree}, non-generic trees or multi-critical branched polymers with $k$ phases  \cite{wheanon}: $$d_s={2k+2 \over 2k+1 }, \quad \mbox{$k=2,3,4\dots$, }$$  random combs $d_s=(4-b)/2, b<2$ \cite{wheacomb}, where $b$ is a power law exponent for the length of the tooth of the random comb, there is no theoretical derivation of the spectral dimension  of random lattices of a given, for instance planar or toroidal, topology. Dynamical triangulations, dual to random lattices, arise as a discretization of the integral over the metrics of some smooth two-dimensional manifold  \cite{dFGZ}. Each triangulation (see Fig. 1) is in one-to-one correspondence with a vacuum diagram of some $N\times N$ Hermitian matrix model. In the large $N$ limit only lattices with planar topology survive  \cite{BIPZ,IZ}.

We show that the random walk is exactly equivalent to  the $q$-state Potts model with non-zero magnetic field taking in the limit $q\to 0$ \cite{Wu}. On a random lattice this model  is defined by the partition function:

\beq\label{pfq} Z^{(q)}_n(\beta, H) =  \sum_{G_n}\sum_{{\s}} \exp \left( {\beta\over 2} \sum^n_{i,j=1} G^{(n)}_{ij} \dl(\s_i,\s_j) + Hq^a \sum^n_{i=1}  \dl(1,\s_i) \right), \eeq
where $G^{(n)}_{ij}$ is the adjacency matrix of the graph, the upper index $n$ is the number of vertices in the graph, indices $i,j=1\dots n$ enumerate vertices, spin variables $\s_i$ associated with the vertex $i$ take $q$ different values (colours) enumerating independent components of the spin, $\sum_{G_n}$ represents the sum over all configurations of graphs in the ensemble,  $\sum_{\s}$  represents the sum over all configurations of spins, $H$ represents the magnetic field normalized by the temperature, $0<a<1$ is an auxiliary parameter which  is essential in the limit of small $q$, $\beta$ is the product of the inverse temperature  and the coupling constant of spins. The limit $q\to 0$ can be better understood through the cluster representation \cite{Bax} of the model \eref{pfq}.  It also can be defined in terms of the tree-like percolations (spanning forests) on a random graph \cite{Stepht}.  
In this representation the partition function \eref{pfq}  for zero magnetic field $H=0$ is given by \beq Z^{(q)}_n(\beta)= \sum_{G_n}\sum_{trees} B^{b(tree)} ,  \eeq where $\sum_{trees}$ is the sum over all trees  spanning the lattice $G_n$, $b(tree)$ is the number of bonds in a given tree on the lattice $G_n$ and the constant $B$ is connected to $\beta$ in \eref{pfq} through the equality $e^{\beta} = 1+ q^aB$. 

The  behaviour of the return probability, the conditional probability and the square displacement  of the random walk at large times can be found by computing critical exponents of the magnetization $\dl$, the two-point correlation function $\eta$ and the correlation length $\nu$ of the  spin model \eref{pfq} correspondingly \cite{Rudnik}.

The Potts model  on a random lattice \cite{Dq, Eq} belongs to the long list of exactly solvable models which could be reformulated as matrix models. The list includes the Ising model \cite{Ki,KBi},  bond-percolations \cite{Kperc}, tree-percolations \cite{Kq}, the $O(n)$ vector model \cite{koson, Eon}, dilute Potts model \cite{ZJ} and many others \cite{kosdupl, kos}. The limit $q\to 0$ of the Potts model which is relevant for our consideration had been solved by the saddle point technique  \cite{Kq}, by the loop equation technique  \cite{Eq} and, recently, by the combinatorial method \cite{sport}. We generalize this results to the case with non-zero magnetic field which breaks the symmetry of the model. In this scenario there will be two different saddle points. However we shall show how it can be simplified in the limit $q\to 0$.

\bigskip

\epsfbox{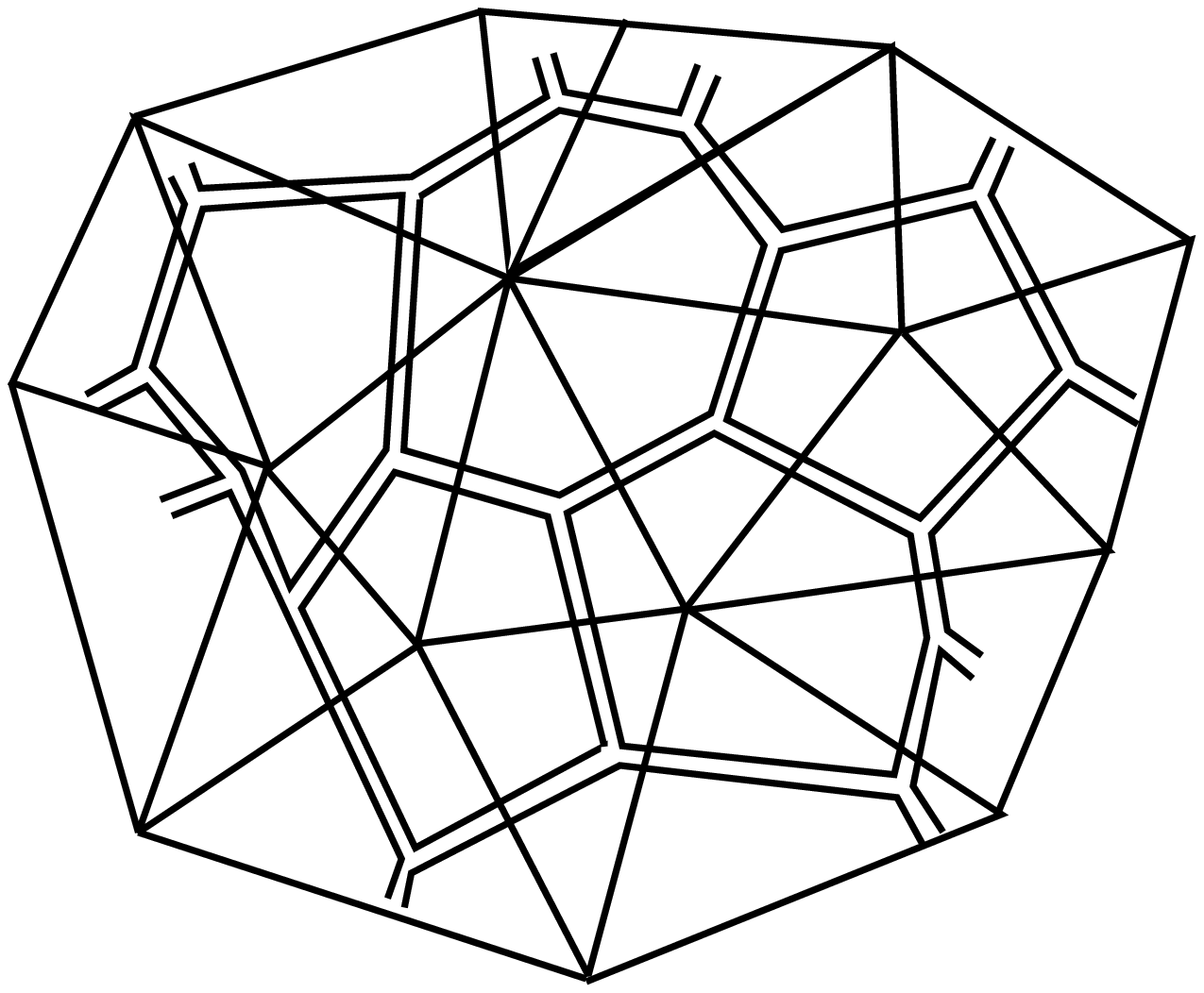}
 
\noindent{{\bf Fig.1}: Duality between fat graphs arising from the perturbation expansion of the $\phi^3$ of one-matrix model and triangulated surface. Each $\phi^3$ vertex of the fat graph  corresponds to the triangle. Gluing together triangles edge to edge is equivalent to the Wick's contraction of $\phi$ matrices.}

\bigskip

We derive the exact  result for the spectral dimension of a random surface and show that in the case of non-zero field there are two phases. One phase corresponds to Hamiltonian walks \cite{duplHam} or branched polymer phase $d_s=4/3$ ($\g_{str}=-1$) another is dilute or pure gravity phase $d_s=2$ ($\g_{str}=-1/2$).

This paper is organized as follows. In section 2 we establish  the equivalence of the random walk and the Potts model. We probe this equivalence computing the critical exponent of the Potts model magnetization on the Bethe lattice and comparing it to the spectral dimension of the corresponding lattice. In section 3 we reformulate the Potts model on a random surface as a random multi-matrix model. In section 4 we find the solution of this model and construct the phase diagram. In section 5 we compare our results with numerical simulations and present our conclusions.

\section{Random walks and the Potts model}

A lattice is a set of $n$ vertices connected by links. It is uniquely defined by the $n\times n$ adjacency matrix $G^{(n)}_{ij}$, whose entries are $G^{(n)}_{ij} =0$ if there is  no link between $i$ and $j$ and $G^{(n)}_{ij}>0$ otherwise. The coordination number or the degree of a vertex is defined as the total number of links connected to it. Consider the ensemble of lattices $G_n$ with $n$ vertices and the random walk jumping on the sites of the such lattices. At each time step the walker with equal probability must jump to the nearest-neighbour site. This process is independent of what can happen to the lattice bonds. We consider that lattice changes its configuration by choosing new one from $G_n$ at  random  every time step. A jump can occur only if the sites are connected by the bonds  at the time the walker attempts to jump. Let $G^{(n)}_{ij}(k)$ denote the adjacency matrix at time step $k$. Suppose that the walker starts at time $k=0$ at the site $0$. Given a particular bond history 
\beq\label{histG} G=G(k) = \{ G^{(n)}_{ij}(0) \dots G^{(n)}_{ij}(k) \} \eeq 
let $p_i(k;G)$ be the probability to find the random walk at the vertex $i$ after time $t$ and $p_i(0) =\dl_{0,i}$. The master equation for  conditional probabilities of the random walk on the dynamical lattice of $n$ sites can be cast into the following system of equations 
\beq\label{nonMark} p_i(k+1;G) = \sum^k_{j=1} \left(G^{(n)}_{ij}(k) p_j(k;G) + [1-G^{(n)}_{ij}(k)] p_i(k;G)\right), \eeq
where the lower case indices $i,j$ enumerate vertices and $k$ counts the number of jumps made by the walker. The random walk is non-Markovian because jumps depend on bond histories $G$. 

It was shown numerically in \cite{fluct} that conditional probabilities $p_i(k;G)$ in the long time limit $k\to +\infty$ did not depend on the particular bond history and approached some average value $\bar p_i(k)$. This universal behaviour is very similar to the behaviour of the spin system on the fluctuating lattice. 

We show  that there is an explicit correspondence between the  random walk and the Potts model. Firstly, we  consider the simplest case when the bond configuration is static. Then conditional probabilities $p_i(k)$ become Markovian because all jumps are independent of earlier events. Then the master equation \eref{nonMark} simplifies  and after taking the continuum time limit it can be written as the system of differential equations 
\beq\label{Mark} \dot p_i = \sum_j G_{ij}  (p_j(t) - p_i(t)), \eeq 
where $G_{ij}$ is the adjacency matrix of some fixed lattice.  After the Laplace transform 
\beq\label{lapl} P_i(H) = \int_0^{+\infty} p_i(t)e^{-Ht}dt \eeq the master equation \eref{Mark} becomes the system of linear equations for the quantities $P_i=P_i(H)$: 
\beq\label{laplME} L_{ij} P_j = \delta_{0,i} \qquad  L_{ij} =H\delta_{ij} - G_{ij}, \eeq 
where $L_{ij} $ is the  Laplacian of a graph with the the adjacency matrix $ G_{ij}$. This system can be easily  solved by inverting the Laplacian. The return probability is defined by the determinant of the Laplacian 
\beq\label{ret} P_0 =  {1\over n} { \pd \ln\det L \over \pd H }.\eeq 
  On the other hand,  it was rigorously proven in \cite{Stephrw} that the determinant of the Laplacian is the sum over all spanning forests on a lattice.  The forest, including $l$ trees which span $m_i$, $i=1\dots l$ vertices correspondingly, gains the weight $H^l \prod_i m_i$, where $\sum_i m_i =n$  and $n$ is the total number of vertices in a lattice. Thus it can be interpreted as the cluster  representation of the partition function of $q\to 0$ Potts model with non-zero magnetic field. The return probability is the magnetization $M$ and it scales as $M\sim H^{1/\dl}$ at the critical point. The spectral dimension is defined by 
  \beq\label{sd} d_s = 2\left(1+{1\over \dl}\right).\eeq 
  
  We consider a generalization of the above result to the case of a random lattice. The key  conjecture is  
   \beq\label{conj} P_0(H)  =  {\pd \over \pd H}\ln \left[\lim_{q\to 0} {Z^{(q)}_n(\beta, H)\over q^{aN}}\right], \eeq  
   where $Z^{(q)}_n(\beta, H)$ is the partition function \eref{pfq}. To see if it is true we consider the graph associated with the path of the walk on a random lattice. The edges of the graph are those bonds on a lattice that the random walk crossed for the first time on its path. It is easy to see that this graph  is a tree which is called the forward tree (see Fig. 2). Thus the partition function of the random walk which is the sum over all possible paths of the random walk is dual to the partition function of all trees on a random lattice \cite{duplrw}.
  
   Our conjecture \eref{conj} is actually a powerful tool for computing exponents of the random walk on different graphs. It allows to employ a critical phenomena technique to the random walk problem which is much broader than methods of a direct solution of the   equation \eref{nonMark}.

 \centerline{\epsfbox{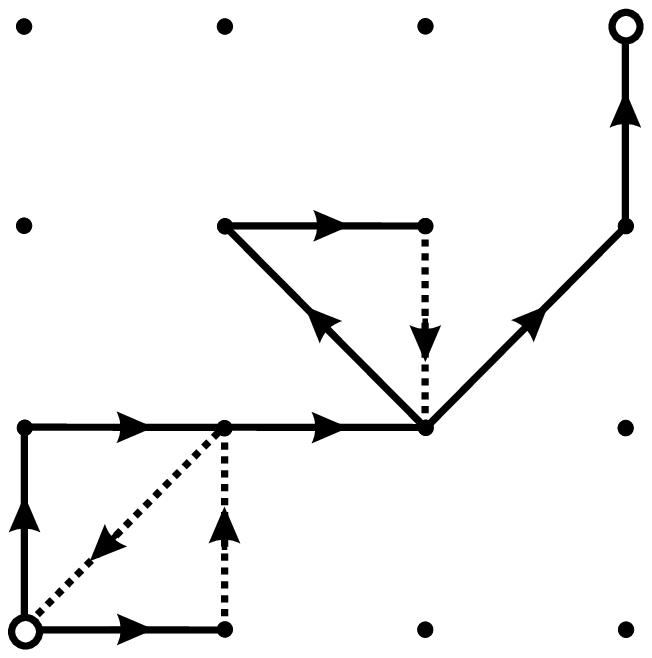}}
\noindent{{\bf Fig.2}: Forward tree is pictured for the path of the random walk on the square regular  lattice.  The walker starts at the lower left corner and ends at the upper right corner. Jumps through the diagonal are also allowed. Arrows denote the direction of jumps. Edges, belonging to the path of the walker, are shown by broken lines. The edge belongs to the forward tree if and only if the walker, jumping through this edge, gets to the vertex it had never visited before. All such edges are shown by solid lines. }

\bigskip

As an example we consider the Bethe lattice with coordination number $z$ (see Fig. 3). We put Potts spins at each site of the lattice.
The solution of this model is well-known \cite{blpm}.  In the thermodynamic limit the magnetization of the central site of the lattice is 
\beq\label{Bmagn} M = \coth[(H-zs)/2], \eeq where $s$ is a parameter defined by $x=e^s$  and $x$ is the fixed point of the recurrence relation which for the case of $q\to 0$ can be written as: 
 \beq\label{Bfp} x= {e^H + (e^{\beta}-2)x^{z-1} \over e^{H+\beta}- x^{z-1} }. \eeq
Recasting  the exponent of the magnetic field from \eref{Bfp}  we find  that 
\beq\label{Bscal} e^H = x^{z-1} {e^{\beta}-2+x \over e^{\beta}x-1}. \eeq
Now we can write $H$ as Taylor series of the small parameter $s$. Up to two leading terms the expansion proceeds  
\beq\label{BHexp} H = (z-2) s + (e^{\beta} -1)^{-1} s^2+\dots .\eeq 
On the other hand from the expansion of the magnetization \eref{Bmagn} one has $M\sim s^{-1}$. Treating $ s$ as a function of $M$ we obtain 
\beq\label{BHexpi} H = (z-2) M^{-1} +(e^{\beta} -1)^{-1} M^{-2}+\dots \eeq
Using the scaling hypothesis $H= M^{\dl} f_s(M^{-1/\beta})$ one has
 \beq\label{Bsd} H = M^{-2} f_s(M).\eeq 
 It gives the values for critical exponents $\dl$ and $\beta$ 
 of the $q$-state Potts model $q\to 0$  on the Bethe lattice
 $\dl =-2$, $\beta =-1$. Knowing $\dl$ and using formula \eref{sd} we derive that $d_s=1$ in agreement with the result \cite{blrw, dorog}.

 \centerline{\epsfbox{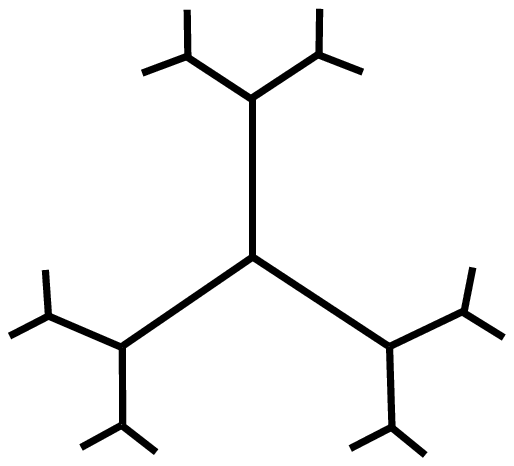}}
\noindent{{\bf Fig.3}: Bethe lattice with coordination number $z=3$.}

\bigskip

\section{Correspondence with the Matrix Model}

From now on we restrict ourselves to the ensemble of random lattices with coordination number 3 and the topology of the sphere. Consider the q-matrix model  defined by the partition function 
\beq\label{MM}\fl Z = \int dM_1\dots dM_q \exp (N\tr [ c \sum^q_{i\ne j}M_iM_j - \sum^q_{i=1} M^2_i + {ge^{Hq^a}\over 3} M_1^3 + \sum^q_{i=2}{g\over 3} M_i^3)  ] ), \eeq
 where $M_i, i=1\dots q$ are  $N\times N$ Hermitian matrices, $H$ is the magnetic field, $0<a<1$ and 
  \beq\label{c} c = 1/(e^{\beta}+q-2),\eeq where  $e^{\beta}-1 = q^aB$. This model generalizes the model of \cite{Eq, Kq} to the case of non-zero magnetic field.  
 
 We note that all matrices in \eref{MM} are coupled to each other. Physically each matrix $M_i$ represents one  component of spin. It can be shown that the propagator is
 \beq\label{prop} \bra \tr M_i M_j \ke_0 = N {c\over (c+1)(1-c(q-1))} \left\{ { (1-c(q-2))/c,   i=j \atop 1,  i\ne j}\right. \eeq where $\bra \dots \ke_0$ denotes the Gaussian average ($g=0$). 
  Using Feynman diagrammatic expansion one would get that the free energy corresponding to \eref{MM}  is equal to the generating function: 
  \beq\label{gf} Z_{T} =\lim_{q\to 0} \sum_{n=1}^{+\infty} \left({cg\over (c+1)(1-c(q-1))}\right)^ne^{Hq^a}Z^{(q)}_n(\beta, H) 
  \eeq 
 where $Z^{(q)}_n(\beta, H)$ is the partition function \eref{pfq}.

 After the change of variables $M_i \to M_i (2(1+c))^{-1/2}$ we have 
 \beq\label{MMrere}\fl Z = \int dM_1\dots dM_q \exp (N\tr [ {h^2\over 2} Y^2 - {1\over 2} \sum^q_{i=1} M^2_i + {\bar ge^{Hq^a}\over 3} M_1^3 + \sum^q_{i=2}{\bar g\over 3} M_i^3)  ] ), \eeq where 
 $\bar g = g(2(1+c))^{-3/2}$, $h^2 =c/(1+c)$ and $Y = M_1+\dots + M_q$.
 By introducing in \eref{MMrere} new Gaussian-distributed random matrix  variable $X$  we replace the first term in the exponent by the matrix polynomial linear in $M_i$: 
 \beq\label{XMM}\fl \int dX\prod_idM_i \exp (N\tr [ -X^2/2  + h X \sum^q_{i=1} M_i  - \sum^q_{i=1} M^2_i + {\bar ge^{Hq^a}\over 3} M_1^3 + \sum^q_{i=2}{\bar g\over 3} M_i^3)  ] ). \eeq

We want to express the integral over matrices \eref{XMM} by the integral over eigenvalues. As it was demonstrated in \cite{GN} the integral over  the matrix in the external field can be  reduced to the integral over eigenvalues by the following formula:
\beq\label{ExtF}\fl \int dM \exp( N\tr [- M^2/2 + MX ] )= \int \prod_{i=1}^N dm_i {\Delta(m)\over \Delta(x)} \exp( N [- m_i^2/2 + m_i x_i ] ) \eeq

 Using \eref{ExtF} and noticing that all  integrals over $M_i$ in \eref{XMM} are  similar, the partition function can be rewritten as:
   \beq\label{XYM} Z = \int \prod^N_{i=1}dx_i \Delta(x)^{2-q} \exp\left(N\sum^N_{i=1} x_i^2/2\right) \Theta_+(x) \Theta_-(x)^{q-1}, \eeq
 where 
 \beq\label{YMp}\fl \Theta_+(x) = \int \prod^N_{i=1} dm_i \Delta(m) \exp \left(N \sum^N_{i=1} [ hx_im_i -  {1\over 2} m_i^2 + {\bar ge^{Hq^a}\over 3} m_i^3]\right),\eeq
  \beq\label{YMm}\fl \Theta_-(x) = \int \prod^N_{i=1} dm_i \Delta(m) \exp \left(N \sum^N_{i=1} [ hx_im_i -  {1\over 2} m_i^2 + {\bar g\over 3} m_i^3]\right).\eeq
  In the present paper we will not give the general solution of \eref{XYM}. For our purposes it is enough to find the partition function when $q\to 0$. In the absence of the magnetic field $H= 0$ functionals $\Theta_+(x)$ and $\Theta_-(x)$ are equal to each other. Hence the value of this functionals will be governed by the same saddle point equation as shown in  \cite{Kperc, Kq}. By noticing that 
  $h^2 = 1/q^aB$ and after the change of variables $m_i\to m_i/\sqrt{q^aB}$ one has
 \beq\label{YMpzero} \Theta_+(x) = \int dm_i \Delta(m)\exp ({N\over q^a B}\sum^N_{i=1} [ x_im_i -  {1\over 2} m_i^2 - {Ge^{Hq^a}\over 3} m_i^3]),\eeq
   where $G= g (\sqrt{2} q^aB)^{-3}$. When $q$ is small the prefactor in the exponent becomes large and the steepest descent method can be used to compute \eref{YMpzero}. Unlike the usual large-N limit the contribution from the  Van-der-Monde determinant will be small and can be neglected. To the leading order we have 
  \beq\label{YMpSD} \Theta_+(x) =  \exp \left({N\over q^a B}\sum^N_{i=1} [ x_iu_i -  {1\over 2} u_i^2 - {Ge^{Hq^a}\over 3} u_i^3]\right)\eeq
  Similarly we have 
  \beq\label{YMmSD} \Theta_-(x) =  \exp \left({N\over q^a B}\sum^N_{i=1} [ x_iv_i -  {1\over 2} v_i^2 - {G\over 3} v_i^3]\right)\eeq
   where $u_i$ and $v_i$ are given by the saddle point condition:
   \beq\label{sadlpt} x_i=u_i+Ge^{Hq^a} u_i^2 \quad  x_i=v_i+G v_i^{2}.\eeq
   In the limit of small $q$ one can express the solution $u_i$ in terms of $v_i$ as perturbation series. Choosing the ansatz 
   \beq\label{anzu}u_i = v_i + \ep v^{(1)}_i \quad \ep=Hq^a\eeq and expanding the exponent $e^{Hq^a}$  we have 
   \beq\label{solu} u_i = v_i - \ep {v^2_i\over 1+2v_i}\eeq
   By doing this the ratio of \eref{YMpSD} and \eref{YMmSD} is significantly simplified: 
   \beq\label{pom} {\Theta_+(x) \over \Theta_-(x)} = \exp\left(-{NGH \over 3B}\sum^N_{i=1}v_i^3 \right)\eeq
 It follows that the partition function is
 \beq\label{XYMq} Z = \int \prod^N_{i=1}dx_i \Delta(x)^{2} \exp\left(N\sum^N_{i=1} x_i^2/2\right) \exp\left(-{NGH \over 3B}\sum^N_{i=1}v_i^3 \right)\eeq
 After changing variables from $x$ to $v$ and using \eref{sadlpt} the integral \eref{XYMq} becomes   
 \beq\label{Xsd} Z = \int dv_i \prod_{i,j}(1+G(v_i+v_j))\Delta(v)^2 \exp [-N\sum_i V(v_i)],\eeq
 \beq V(x)= \left( {1\over 2}(x+Gx^2)^2+   {HG x^3 \over 3B} \right).\eeq

 Again, we apply the steepest descent method but now with respect to large $N$. The saddle point equation is
 \beq\label{eigv} 2\sum_{j\ne i} {1\over v_i - v_j} + \sum_{j} {2G \over 1+ G(v_i + v_j)} = N V'(v_i). \eeq
 The distribution of the eigenvalues becomes continuous  with density $\rho(x) = (1/N) \sum \dl(x-v_i)$. We restrict ourselves to the one-cut case where all eigenvalues $v_i$ belong to the support  consisting of one interval $[a,b]$, $ab>0$. 
The equation for the density is the integral equation:
\beq\label{evint} \mbox{P} \int {\rho(y)dy \over x-y} + \int {\rho(y)dy \over 1/G + x+y} = {1\over 2} V'(x),\eeq
where P denotes the principal value of the integral. Introducing the trace of the resolvent \beq\label{res} \om(x) = {1\over N} \tr {1\over M-x } = \int {\rho(y)dy \over y-x} \eeq 
\eref{evint} can be equivalently written as
\beq\label{reseq} \om(x+i0) + \om(x-i0) + 2 \om(-x-1/G) = {1\over 2} V'(x)\eeq
The integral equation that governs the eigenvalue density is 
 \beq\label{evd}\int {dy \rho(y) \over (x-y)(1+G(x+y)) } = {1\over 2}(1+ Gx)x + {HG\over 2B }{x^2 \over 1+ 2Gx}\eeq

\section{Solution}

The equation \eref{evd} represents  the Riemann-Hilbert problem. The solution can be found \cite{gakh}: 

\beq\label{rhosol}\fl \rho(x-1/2G) ={1\over \pi} [(x^2-a^2)(b^2-x^2)]^{1/2} \int_a^b {f(y)dy \over [(y^2-a^2)(b^2-y^2)]^{1/2} (x^2-y^2)}, \eeq
where 
\beq\label{fofx} f(x) = Gx^3 + HGx^2/(2B) -(1/4+H/(2B))x +H/(8BG),\eeq 
supplied with additional condition: 
\beq\label{rhocond}\int_a^b {f(y)dy \over [(y^2-a^2)(b^2-y^2)]^{1/2}} =0 \eeq
and with the normalization condition:
\beq\label{rhonorm}\int_a^b dx \rho(x) =1. \eeq
The ends of the interval $[a,b]$ are functions of $G,B,H$, to be determined from transcendental equations  \eref{rhocond} and \eref{rhonorm}   on $a=a(G,B,H)$ and $b=b(G,B,H)$.

We will need the first equation which can be resolved as

\beq\label{soleli}\fl G{\pi\over 2} \left(1 +{1\over 2} (b^2 - a^2)\right) + {HGa\over 2B} E\left(1-{b^2\over a^2}\right) - \left({1\over 4}+{H\over 2B}\right){\pi\over 2} + {H\over 8BGa}K\left(1-{b^2\over a^2}\right)=0,\eeq
where $K(x),E(x)$  are elliptic integrals of the first and second kind respectively. We used the following integrals: 
\beq\label{intone}  \int_a^b{dy \over [(y^2-a^2)(b^2-y^2)]^{1/2}} =a^{-1}K\left(1-{b^2\over a^2}\right),\eeq
\beq\label{inttwo} \int_a^b{ydy \over [(y^2-a^2)(b^2-y^2)]^{1/2}} ={\pi\over 2},\eeq
\beq\label{intthr} \int_a^b{y^2dy \over [(y^2-a^2)(b^2-y^2)]^{1/2}} =aE\left(1-{b^2\over a^2}\right),\eeq \beq\label{intfour} \int_a^b{y^3dy \over [(y^2-a^2)(b^2-y^2)]^{1/2}} = {\pi\over 2} \left(1 +{1\over 2} (b^2 - a^2)\right)\eeq

Then the critical behaviour  of the partition function \eref{MM} is obtained by taking the double scaling limit $B \to B_c$ (infinite random surface) and $G \to G_c $ (infinite trees). The latter occurs  near the upper edge $b$ of the support. We have $b_c=-1/2G_c$. It can be recast in the following form 

\beq\label{omega} w=1\pm 1/h, h>1 \qquad  1 =  \cos(\pi/h) \quad h=2,4\dots \eeq Then the critical density is 
\beq\label{rhocrit} \rho(x)\sim (g-g_*)(b-x)^{1-1/h} + (b-x)^{1+1/h} \eeq and string susceptibility is $\g_{str} = -1/h$. The solution is singular over $[a-1/2G,b-1/2G]$ and $[-b-1/2G,-a-1/2G]$. Branching points coincide when $a=0$ \cite{Kq}. It can be shown that the Potts model is in the critical point if this condition is satisfied. Thus there are two phases of the model: dilute and dense. In the dilute phase the problem is equivalent to the (-2)-dimensional dynamical triangulated surface. The critical exponent of the string susceptibility is $\g_{str} =-1$. In the dense phase defined by the condition $b=-1/2G$ one has $\g_{str} =-1/2$. From the equation \eref{soleli} we can see that 
\beq\label{bH} { \pd b \over \pd H} \sim {1 \over  H}.\eeq

 Finally, let us compute the magnetization $M = - \pd F /\pd H $. The free energy is 
 \beq\label{pfsol}\fl F = \int_a^b dx\rho(x) V(x) + \int_a^b \int_a^b dx dy \rho(x)\rho(y)[\ln|x-y|+ \ln(1+G(x+y)) ]. \eeq It is linear in $H$ for the small magnetic field. Taking the derivative one can obtain that 
 \beq\label{M} M\sim {\pd b \over \pd H} (C_1 H + C_2 H^2) = (C_1  + C_2 H) = f_s(H).\eeq
 It means that $1/\dl = 0$. Substituting the exponent $\dl$ in \eref{sd} one has 
 \beq\label{sdrs} d_s=2. \eeq

\section{Conclusion}

In conclusion, we have demonstrated an alternative derivation of the spectral dimension of the random surface. We note that our results match the numerical simulations \cite{ambj}. It should also coincide with the KPZ \cite{KPZ} result for the conformal field theory with central charge $c=-2$ coupled to the two-dimensional gravity. The result can be generalized to  topologies with higher genus by the DDK \cite{D, DK} formula. We notice that it should not change the value of the spectral dimension $d_s$. The only sensitive exponent to DDK is the string susceptibility $\g_{str}$.

\section*{Acknowledgment}
 
 I would like to thank Dmitry Krotov and Sergei Alexandrov for valuable discussions.  I would like to acknowledge support from Ajay Gopinathan via his start-up funds and his James S. McDonnell Foundation Award.

\end{document}